\documentclass[preprint,onecolumn,nofootinbib]{revtex4}
\usepackage[colorlinks=true,linkcolor=blue,urlcolor=blue,filecolor=black,citecolor=red,pdfstartview=FitV,pdftitle={},pdfsubject={},pdfkeywords={},pdfpagemode=None,bookmarksopen=true]{hyperref}
\usepackage{graphicx}
\usepackage{amsmath}
\usepackage{amsfonts}
\usepackage{amssymb,ulem}
\usepackage{color,xcolor}%
\usepackage{CJK}
\usepackage{subfigure}
\usepackage{amsthm,amsmath,amssymb}
\usepackage{mathrsfs}
\usepackage{url}
\usepackage{hyperref}
\usepackage{array}

\usepackage{dcolumn}
\usepackage{float}
\usepackage{appendix}
\begin{document}
\title{Quasinormal modes of quantum-corrected black holes}

\author{Huajie Gong$^{1}$}
\thanks{huajiegong@qq.com}
\author{Shulan Li$^{1}$}
\thanks{shulanli.yzu@gmail.com}
\author{Dan Zhang$^{2}$}
\thanks{danzhanglnk@163.com}
\author{Guoyang Fu$^{3}$}
\thanks{FuguoyangEDU@163.com}
\author{Jian-Pin Wu$^{1}$}
\thanks{jianpinwu@yzu.edu.cn, corresponding author}
\affiliation{
	$^1$ Center for Gravitation and Cosmology, College of Physical Science and Technology, Yangzhou University, Yangzhou 225009, China}
\affiliation{
	$^2$ Key Laboratory of Low Dimensional Quantum Structures and Quantum Control of Ministry of Education, Synergetic Innovation Center for Quantum Effects and Applications, and Department of Physics, Hunan Normal University, Changsha 410081, Hunan, China}
\affiliation{
	$^3$ Department of Physics and Astronomy,
	Shanghai Jiao Tong University, Shanghai 200240, China}

\begin{abstract}
	
In this paper, we investigate the quasinormal mode (QNM) spectra for scalar perturbation over a quantum-corrected black hole (BH). The fundamental modes of this quantum-corrected BH exhibit two key properties. Firstly, there is a non-monotonic behavior concerning the quantum-corrected parameter for zero multipole number. Secondly, the quantum gravity effects result in slower decay modes. For higher overtones, a significant deviation becomes evident between the quasinormal frequencies (QNFs) of the quantum-corrected and  Schwarzschild BHs. The intervention of quantum gravity corrections induces a significant outburst of overtones. This outburst of these overtones can be attributed to the distinctions near the event horizons between the Schwarzschild and quantum-corrected BHs. Therefore, overtones can serve as a means to probe physical phenomena or disparities in the vicinity of the event horizon.

\end{abstract}

\maketitle
\tableofcontents

\section{Introduction}

General relativity (GR), posited more than a century ago, has successfully withstood all observational tests done up to now. Nevertheless, according to the singularity theorems in GR, the emergence of black holes (BHs) and spacetime singularities is anticipated \cite{PhysRevLett.14.57,hawking1970singularities}. Singularities indicate the failure of GR when the curvature of spacetime becomes infinitely large. Hence, it is reasonable to anticipate that a quantum theory of gravity will emerge and become dominant in these regions, addressing the issue of singularities. Some competing quantum theories of gravity have been proposed thus far. Notably, loop quantum gravity (LQG) stands out as a significant approach that is both background-independent and nonperturbative. It has garnered significant attention and has been extensively studied \cite{Rov,Thiemann:2001gmi,Ashtekar:2004eh,Han:2005km}.

Moreover, the technique of loop quantization has proven successful in quantizing symmetry-reduced cosmological spacetimes, a field known as loop quantum cosmology (LQC) \cite{Bojowald:2001xe, Ashtekar:2006rx, Ashtekar:2006uz, Ashtekar:2006wn, Ashtekar:2003hd, Bojowald:2005epg, Ashtekar:2011ni, Wilson-Ewing:2016yan}. Effective LQC theory can be constructed by incorporating two fundamental quantum gravity effects, namely the inverse volume correction and the holonomy correction. The quantum gravity effects in LQC establish connections with low-energy physics, thereby furnishing a tractable cosmological framework to explore and comprehend quantum gravity events. Remarkably, these effects effectively evade the big bang singularity that is inherent in classical GR \cite{Bojowald:2001xe, Ashtekar:2006rx, Ashtekar:2006uz, Ashtekar:2006wn, Ashtekar:2003hd, Bojowald:2005epg, Ashtekar:2011ni, Wilson-Ewing:2016yan, Bojowald:2003xf, Singh:2003au, Vereshchagin:2004uc, Date:2005nn, Date:2004fj, Goswami:2005fu, Papanikolaou:2023crz}, and instead replace it with a non-singular big bounce, even at the semi-classical level \cite{Bojowald:2005zk, Stachowiak:2006uh}. Significant progress has been made in recent years in the application of LQC to the early universe. For comprehensive examples, please refer to \cite{Bojowald:2001xe,Ashtekar:2006rx,Ashtekar:2006uz,Bojowald:2002nz,Bojowald:2003mc,Tsujikawa:2003vr,Wu:2012mh,Wu:2018mhg,Wu:2010wj,Li:2023axl,Han:2017wmt,Renevey:2021tmh,DeSousa:2022rep} and references therein.

Following the analogous concept in LQC \cite{Bojowald:2001xe, Ashtekar:2006rx, Ashtekar:2006uz, Ashtekar:2006wn, Ashtekar:2003hd, Bojowald:2005epg, Ashtekar:2011ni, Wilson-Ewing:2016yan}, various effective models of BHs incorporating corrections from LQG have been developed. Noteworthy instances of these models are documented in \cite{Ashtekar:2005qt, Modesto:2005zm, Modesto:2008im, Modesto:2009ve, Campiglia:2007pr, Bojowald:2016itl, Boehmer:2007ket, Chiou:2008nm, Chiou:2008eg, Joe:2014tca, Yang:2022btw, Lewandowski:2022zce, Gan:2022oiy, Vagnozzi:2022moj, Afrin:2022ztr}, along with pertinent references. A distinctive characteristic of LQG BHs is the replacement of the singularity by a transition surface connecting a trapped region to an anti-trapped region, reminiscent of an inner region of a BH and a white hole.

Recently, a novel quantum BH has been proposed by exploring the quantum Oppenheimer-Snyder (qOS) model within LQC \cite{Lewandowski:2022zce}. The quantum effects establish a minimum threshold for the mass of the BH formed from the collapsing dust ball. In scenarios involving larger masses, an event horizon forms during the collapse. Quantum gravitational effects intervene, halting the collapse of the dust matter as the energy density approaches the Planck scale, subsequently triggering a bounce into an expanding phase. This resolution effectively addresses the singularity inherent in classical BHs. 

Subsequently, the properties of BH shadows and images have been investigated in recent studies \cite{Yang:2022btw,Zhang:2023okw}. Notably, quantum gravitational corrections have been found to impact the sizes of BH shadows \cite{Yang:2022btw}. Additionally, it has been demonstrated in \cite{Zhang:2023okw} that radiations emitted by a BH companion can traverse its horizon, travel into the deep Planck region, and ultimately manifest from a white hole in our universe. The observed BH image exhibits additional luminous rings caused by these radiations, with some of these rings distinctly visible in the shadow region. Consequently, the BH image contains encoded information about quantum gravity. Furthermore, the researchers in \cite{Yang:2022btw} have investigated the stability of the quantum-corrected BH by computing QNMs. The findings demonstrate that the quantum-corrected BH remains stable when subjected to scalar and vector perturbations. 

In this study, we take a significant step forward by investigating not only the fundamental mode but also several high overtone modes of the scalar perturbation in the context of the LQG-corrected BH. Recent studies \cite{Giesler:2019uxc,Oshita:2021iyn,Forteza:2021wfq,Oshita:2022pkc} have revealed that in order to accurately model the ringdown phase, it is imperative to account for the contribution of the first few overtones, contrary to the prevailing belief that the fundamental mode predominantly governs the signal. Additionally, there is noteworthy evidence suggesting significant excitation of overtones in specific scenarios, and the possibility of detecting such excitations using space-based gravitational wave (GW) detectors like LISA during the initial ringdown phase has been proposed \cite{Oshita:2022yry}.

Furthermore, a recent study conducted by Konoplya et al. has unveiled a connection between the characteristics of the high overtones and the the geometric structure around the event horizon of a BH \cite{Konoplya:2022pbc}. If alterations to Einstein's theory predominantly affect the geometric properties of BHs in close proximity to the event horizon, the fundamental mode often remains essentially unaltered. Nevertheless, even a minor alteration in the vicinity of the event horizon may have a significant impact on the first few overtones \cite{Konoplya:2022pbc}, thus providing an avenue to investigate the event horizon structure.
In addition, when compared to echoes, overtones exhibit a considerably higher energy contribution, making them a crucial area of exploration for the study of various BHs \cite{Konoplya:2022hll,Konoplya:2022iyn,Konoplya:2023aph}. These notable findings provide further motivation for our continued investigation into the overtones of distinct BHs. With the above motivations in mind, we will delve into the study of the characteristics of the high overtones of the scalar perturbation of the LQG-corrected BH.

The paper is organized as follows. In Section \ref{LQG-model}, we will provide a brief overview of the scalar perturbation of the quantum-corrected BH. Section \ref{qnms-scalar} explores QNMs, encompassing both the fundamental mode and the emergence of overtones. Section \ref{sec-conclusion} presents the conclusions and discussions. The numerical calculation methods and error analysis of QNMs are detailed in Appendix \ref{sec-methods}.

\section{Scalar field over the quantum-corrected BH}\label{LQG-model}

Recently, a novel quantum BH has been introduced by studying the qOS model within LQC \cite{Lewandowski:2022zce}. The maximal extension structure of this spacetime closely resembles that of the Reissner-Nordström (RN) BH spacetime. It features two Killing horizons, with the outer horizon corresponding to the event horizon of a black/white hole, and the inner horizon corresponding to the Cauchy horizon. However, akin to the scenario in the RN spacetime, the Cauchy horizon may experience instability under perturbations. The region outside the outer horizon is static and asymptotically flat. In contrast, the region between the two horizons is not static and constitutes the trapped/anti-trapped region. The region within the inner horizon is static, linking the trapped and anti-trapped regions. Notably, unlike the RN spacetime, there exists a non-zero bounce radius $r=r_b$ beyond the classical timelike singularity at $r=0$, where the region $r<r_b$ is deemed physically inaccessible. Note that the above discussion primarily applies to scenarios where the mass of the BH is greater than the minimum mass. In the event that the BH mass equals the minimum mass, it exhibits characteristics akin to an extremal BH, resembling the RN case, where the two horizons coincide.

\begin{figure}[htbp]
	\centering
	\includegraphics[width=0.6\textwidth]{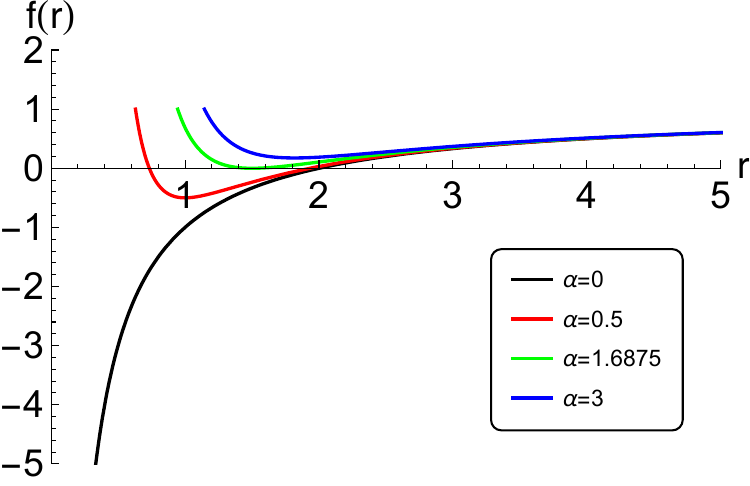}\vspace{0.4mm}
	\caption{The BH factor $f(r)$ for various $\alpha$.}
	\label{fvsr}
\end{figure}

In our study, we exclusively concentrate on the exterior spacetime of this quantum-corrected model, described as \cite{Lewandowski:2022zce}\footnote{The spacetime metric \eqref{metric}, serving as an effective quantum-corrected Schwarzschild geometry, is consistent with metrics derived in prior works \cite{Kelly:2020uwj, Parvizi:2021ekr, Husain:2022gwp, Giesel:2022rxi, Jusufi:2022uhk}. In addition, by employing effective field theory techniques to general relativity, one can also obtain quantum-corrected BH solutions. Examples of such solutions can be found in literature, such as \cite{Calmet:2021lny,Xiao:2021zly,Battista:2023iyu}.}
\begin{eqnarray}
	&&
	ds^2=-f(r)dt^2+f(r)^{-1}dr^2+r^2 d\theta^2+r^2 \sin^2{\theta} d\phi\,,\nonumber
	\
	\\
	&&
	f(r)=1-\frac{2M}{r}+\frac{\hat{\alpha} M^2}{r^4}\,,
	\label{metric}
\end{eqnarray}
where $\hat{\alpha} = 16 \sqrt{3} \pi \gamma^3 {l_p}^2$ represents the quantum-corrected parameter, with $\gamma$ being the Immirzi parameter and $l_p=\sqrt{\hbar}$ denoting the Planck length, and $M$ stands for the mass of the BH. Typically, the Immirzi parameter is considered a free parameter, and thus, we treat $\hat{\alpha}$ as a free variable in this paper. For convenience, we introduce the dimensionless parameter $\alpha =\hat{\alpha}/M^{2}$. Throughout this study, we adopt the convention $M=1$. For the BH spacetime to possess an event horizon, the quantum-corrected parameter should satisfy the following constraint:
\begin{eqnarray}
0\leq\alpha\leq\frac{27}{16}\,.
	\label{constrain-alpha}
\end{eqnarray} 
When $\alpha=0$, the above quantum-corrected geometry reduces to the Schwarzschild case. In particular, for $0<\alpha<27/16$, the quantum-corrected geometry \eqref{metric} exhibits two horizons. Conversely, when $\alpha$ reaches the upper bound, i.e., $\alpha =27/16$, the geometry possesses a degenerate horizon. To visually represent this, we display the redshift factor $f(r)$ for various $\alpha$ in Fig. \ref{fvsr}.

Then, we investigate the perturbation of the massless scalar field $\Phi$ over this quantum-corrected BH and explore their response. The general covariant Klein-Gordon (KG) equation for a scalar field is given by:
\begin{eqnarray}\label{scalar-equation}
	\frac{1}{\sqrt{-g}}\partial_\nu(g^{\mu\nu}\sqrt{-g}\partial_\mu\Phi)=0,
\end{eqnarray}
where $g_{\mu\nu}$ is the background metric. Considering the spherical symmetry of the spacetime, we can separate the perturbation field $\Psi$ into the following form:
\begin{eqnarray}\label{separate}
	\Phi(t,r,\theta,\phi)=Y_{l,m}(\theta,\phi)\frac{\Psi(t,r)}{r},
\end{eqnarray}
where $Y_{l,m}(\theta,\phi)$ is the spherical harmonics, with $l$ and $m$ being the multipole and azimuthal quantum number, respectively. Then, we can transform Eq. (\ref{scalar-equation}) into the following wave-like equation:
 \begin{eqnarray}\label{wavelike-equation}
 	-\frac{\partial^{2}\Psi}{\partial t^2}+\frac{\partial^{2}\Psi}{\partial r^{2}_*}-V\Psi=0,
 \end{eqnarray}
where $dr_*=dr/f(r)$, and the effective potential $V$ is defined as:
 \begin{eqnarray}\label{effective-potential}
 	V=f \left( \frac{l(l+1)}{r^2}+\frac{f'}{r}\right) .
\end{eqnarray}

\begin{figure}[htbp]
	\centering
	\includegraphics[width=0.48\textwidth]{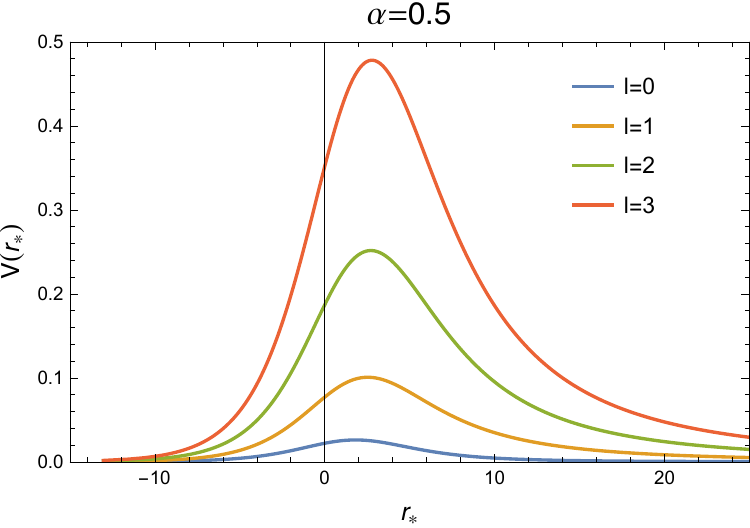}\hspace{4mm}
	\includegraphics[width=0.48\textwidth]{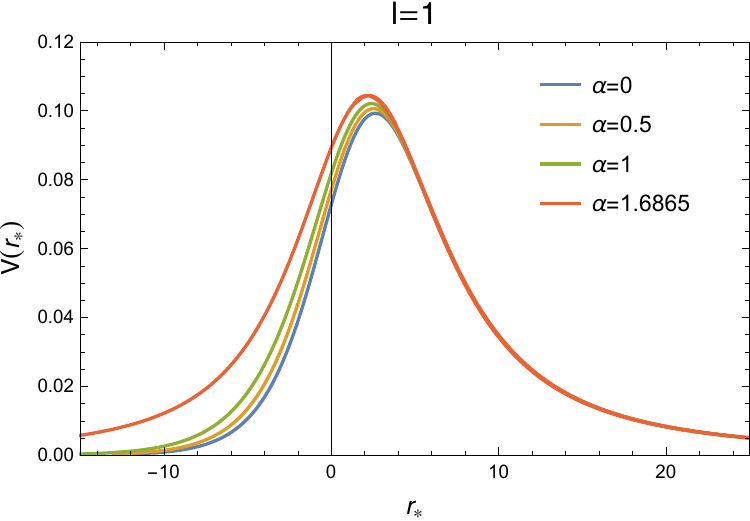}\vspace{0.4mm}
	\caption{Left plot: The effective potential $V(r_*)$ for different $l$ with $\alpha=0.5$. Right plot: The effective potential $V(r_*)$ for different $\alpha$ with $l=1$.}
	\label{Vvsrsv1}
\end{figure}

Fig.\ref{Vvsrsv1} illustrates the effective potential $V(r_*)$ for various values of $\alpha$ with $l=1$. As shown in Fig.\ref{Vvsrsv1}, when the quantum parameter $\alpha$ is fixed, the height of $V(r_*)$ increases rapidly with the multipole number $l$. This behavior aligns with what is observed in Schwarzschild or RN BHs. However, a subtle deformation in the effective potential near the horizon is noticeable when varying the quantum-corrected parameter $\alpha$. In the subsequent sections, we will explore the implications of this slight deformation in the near-horizon region on the QNMs.

\section{Quasinormal modes}\label{qnms-scalar}

In this section, we are prepared to investigate the properties of the QNMs over the LQG-corrected background \eqref{metric}. To work in the frequency domain, we expand $\Psi$ as $\Psi=e^{-i\omega t}\psi$. Consequently, the KG equation takes the form:
\begin{eqnarray}\label{scalar-eqution2}
	\frac{\partial^2 \psi}{\partial r_*^2}+(\omega^2-V)\psi=0\,.
\end{eqnarray}

The process of determining the QNMs involves solving an eigenvalue problem. For this purpose, we will impose the following boundary conditions:
\begin{eqnarray}\label{bdy}
	\psi \sim e^{\pm i \omega r_*}\,,~~~~~~ \;r_*\rightarrow \pm \infty\,,
\end{eqnarray}
These boundary conditions indicate that the waves are purely outgoing at infinity and purely ingoing on the event horizon, prohibiting any waves from propagating outward from the horizon or inward from infinity. Such conditions characterize a BH's response to a transient perturbation, occurring after the source has ceased to exert an influence \cite{Konoplya:2011qq, Berti:2009kk, Kokkotas:1999bd}.

There are numerous methods available for calculating quasinormal frequencies (QNFs). In this study, we provide a brief overview of three methods: the WKB method, the pseudospectral method, and the Asymptotic Iteration Method (AIM), which are discussed in detail in Appendix \ref{sec-methods}. Additionally, we perform an error analysis of these methods while computing the QNFs of the quantum-corrected BH, as detailed in the same appendix. The main focus of this section is on the presentation and discussion of the physical results.

\subsection{Fundamental modes}

In \cite{Yang:2022btw}, the authors have worked out the fundamental modes for $l=2$ and discovered that this LQG-corrected BH is stable under the scalar and vector perturbations. In this subsection, we will delve further into investigating the general properties of the fundamental modes resulting from the LQG correction.
\begin{figure}[htbp]
	\centering
	\includegraphics[width=0.45\textwidth]{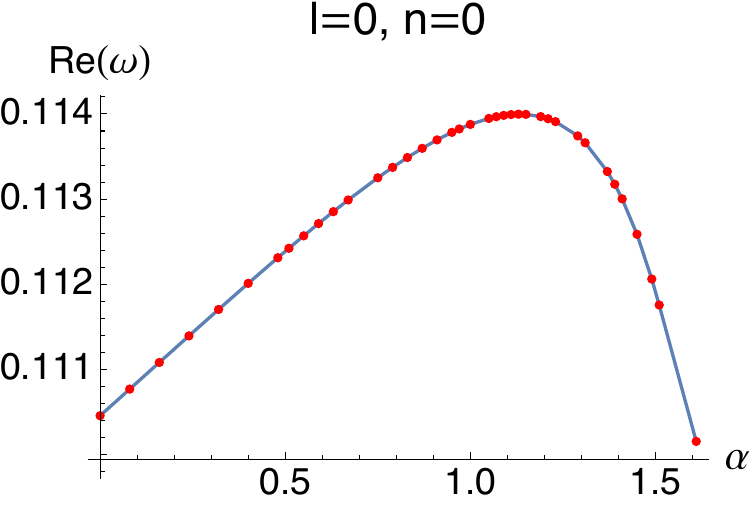}\hspace{4mm}
	\includegraphics[width=0.45\textwidth]{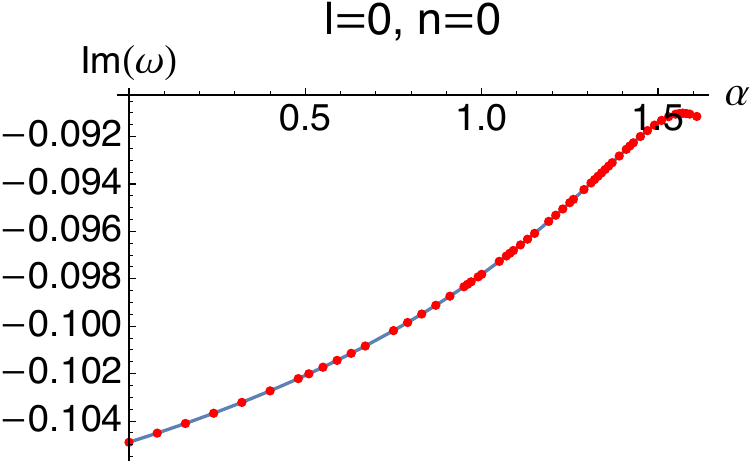}
	\caption{QNFs as a function of the LQG-corrected parameter $\alpha$.}
	\label{ReImn0}
\end{figure}

Fig.\ref{ReImn0} illustrates the real and imaginary parts of the fundamental modes as functions of the LQG-corrected parameter $\alpha$ for $l=0$ and $n=0$. We observe that the real part of the QNFs, $Re(\omega)$ exhibits a pronounced non-monotonic behavior with respect to the LQG-corrected parameter $\alpha$. Specifically, as $\alpha$ increases, indicating a greater deviation from the Schwarzschild BH, $Re(\omega)$ initially increases, signifying a stronger oscillation of the system, and then decreases. It is worth noting that as the parameter $\alpha$ approaches a value close to extremity, $Re(\omega)$ becomes less than the value of the Schwarzschild BH for the same parameter. At the same time, with an increase in $\alpha$, the imaginary part of the QNFs, $Im(\omega)$ generally rises, indicating a smaller damping rate. Only when $\alpha$ approaches the near-extremal value does $Im(\omega)$ diminish slightly. Overall, however, $Im(\omega)$ remains greater than that of the Schwarzschild BH. This observation implies that the presence of quantum gravity effects results in slower decay modes.

\begin{figure}[htbp]
	\centering
	\includegraphics[width=0.45\textwidth]{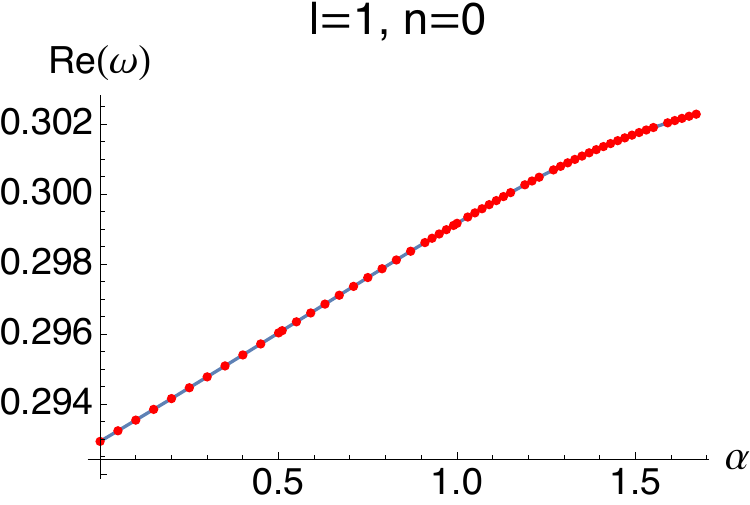}\hspace{0.8mm}		
	\includegraphics[width=0.45\textwidth]{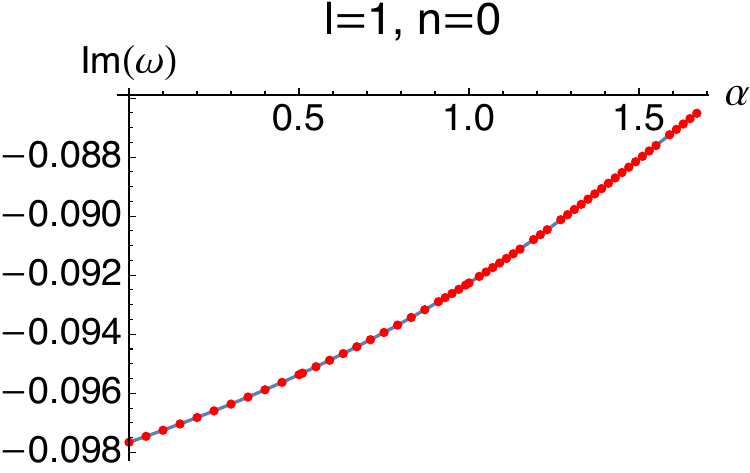}		
	\caption{QNFs as a function of $\alpha$ for the massless scalar field perturbation.}
	\label{ReImn0v1}
\end{figure}

Then, we proceed to calculate the QNFs as a function of $\alpha$ for $l=1$ and $n=0$ in Fig.\ref{ReImn0v1}. We observe that both the real and imaginary parts show a clear monotonic behavior with respect to $\alpha$, which differs from the case of $l=0$. Specifically, as $\alpha$ increases, both $Re(\omega)$ and $Im(\omega)$ monotonically increase, indicating that the system oscillates stronger and decays slower. Furthermore, we also examine the fundamental mode for $l>1$ and arrive at a similar conclusion as in the case of $l=1$.

\subsection{Higher overtones}\label{sec-highovertones}

\begin{figure}[htbp]
	\centering
	\includegraphics[width=0.45\textwidth]{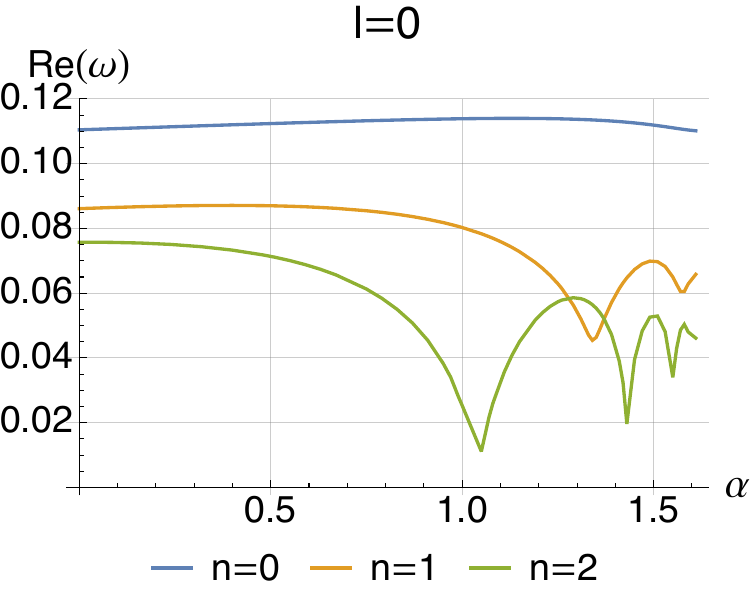}\hspace{4mm}	
	\includegraphics[width=0.45\textwidth]{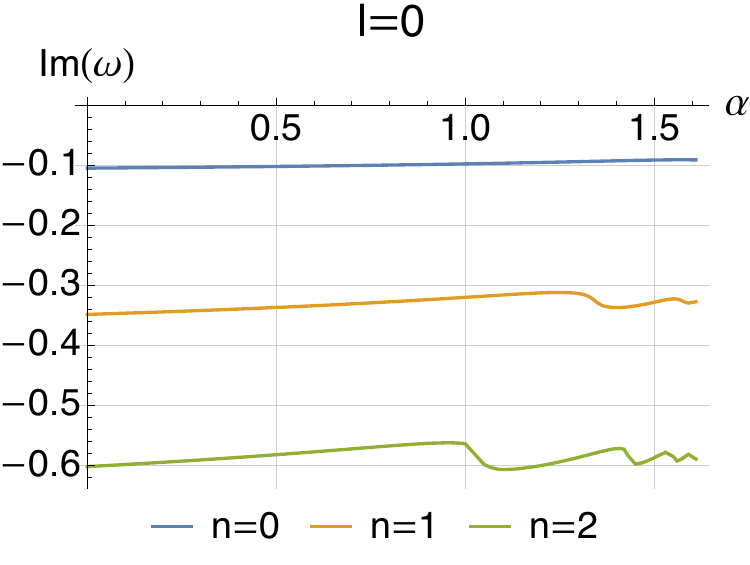}
	\caption{$Re(\omega)$ (left) and $Im(\omega)$ (right) as a function of $\alpha$ for the fundamental mode and first two overtones from top to bottom.}
	\label{largenl0}
\end{figure}

\begin{figure}[htbp]
	\centering
	\includegraphics[width=0.45\textwidth]{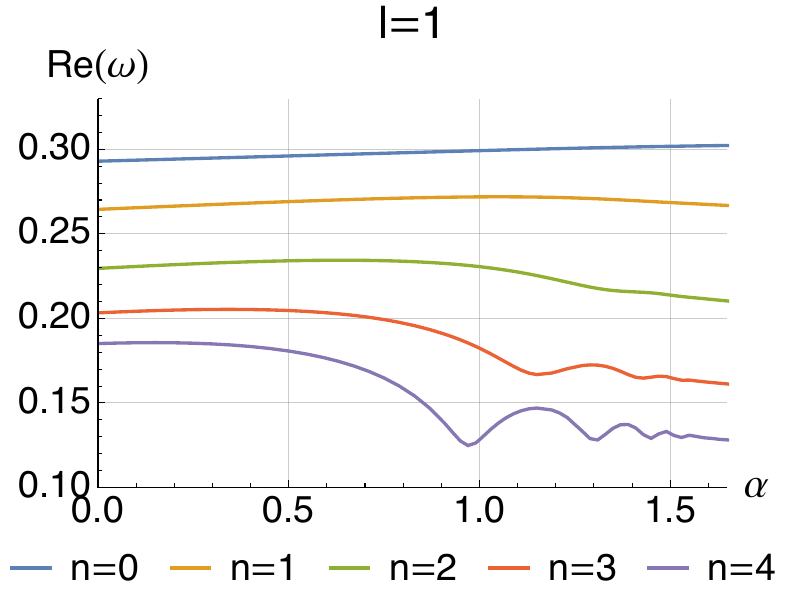}\hspace{4mm}	
	\includegraphics[width=0.45\textwidth]{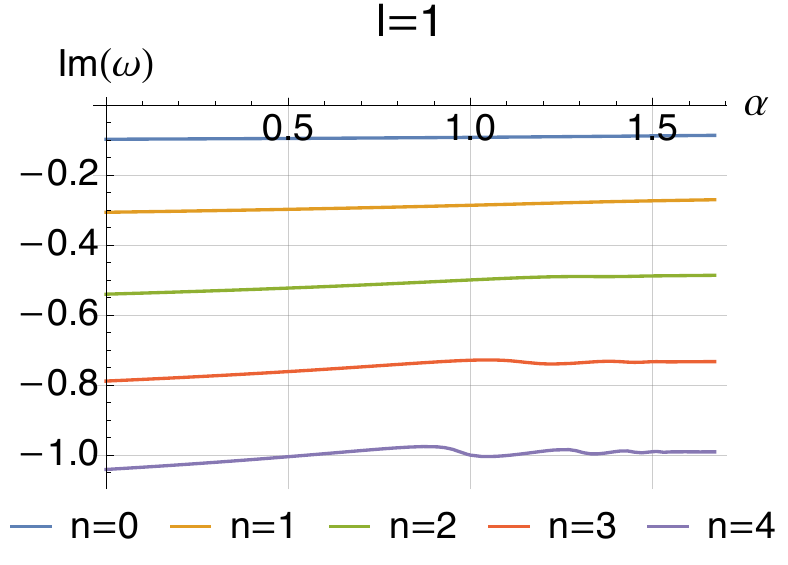}
	\caption{$Re(\omega)$ (left) and $Im(\omega)$ (right) as a function of $\alpha$ for the fundamental mode and first four overtones from top to bottom.}
	\label{largenl1}
\end{figure}

As elucidated in the preceding subsection, the fundamental modes display intriguing properties due to the quantum gravity effect. However, upon closer analysis, it becomes evident that the changes in the fundamental modes' absolute values are exceedingly minor, as evidenced in Table \ref{ta-1}. The table demonstrates that for the fundamental modes, a subtle disparity between the values of the QNFs of the LQG-corrected BH and of the Schwarzschild BH is only noticeable at the third decimal place. This discrepancy is also illustrated graphically by the blue curves corresponding to $n=0$ in Fig.\ref{largenl0} and Fig.\ref{largenl1}, aiding in visual comprehension.
\begin{table}[htbp]
	\centering
	\setlength\tabcolsep{10pt}
	\caption{The QNM spectra for the massless scalar field perturbation with different $n$, $l$ and $\alpha$.}
	\begin{tabular}{|c|c|c|c|c|c|c|c|c|c|c|c|c|}
		\hline
		& $l=0$ & $l=1$ \\
		\hline
		$n$ & $\omega(\alpha=0)$  \qquad \, $\omega(\alpha=0.5)$ & $\omega(\alpha=0)$ \,\, \qquad $\omega(\alpha=0.5)$ \\
		\hline
		0 & 0.110455-0.104896$i$ \, 0.112386-0.102076$i$ & 0.292936-0.097660$i$ \, 0.296040-0.095370$i$ \\
		\hline
		1 & 0.086117-0.348052$i$ \, 0.087003-0.336512$i$ & 0.264449-0.306257$i$ \, 0.269031-0.297772$i$ \\
		\hline
		2 & 0.075744-0.601084$i$ \, 0.071330-0.581330$i$ & 0.229539-0.540133$i$ \, 0.234050-0.522719$i$ \\
		\hline
		3 & 0.070451-0.853744$i$ \, 0.057201-0.827698$i$ & 0.203258-0.788298$i$ \, 0.204636-0.761206$i$ \\
		\hline
		4 & 0.067386-1.105623$i$ \, 0.041539-1.071711$i$ & 0.185109-1.040762$i$ \, 0.180695-1.004284$i$ \\
		\hline
		5 & 0.064180-1.357729$i$ \, 0.035752-2.136065$i$ & 0.172076-1.294120$i$ \, 0.159216-1.248703$i$ \\
		\hline
	\end{tabular}
	\label{ta-1}
\end{table}

However, upon investigating the higher overtones, a remarkable deviation becomes apparent between the QNFs of the LQG-corrected BH and the Schwarzschild BH. As shown in Table \ref{ta-1}, it becomes evident that for certain modes of higher overtones, this discrepancy can be substantial, reaching magnitudes of several hundred percent, even when the fundamental modes closely aligns with the Schwarzschild counterpart. This intriguing behavior is vividly illustrated in Fig.\ref{largenl0} and Fig.\ref{largenl1}, providing a clear visual representation of these deviations. Hence, the introduction of quantum gravity corrections triggers this pronounced outburst in the overtones. Similar behaviors have been noted in various geometric settings beyond the Schwarzschild BH, encompassing the RN BH, Bardeen BH, the higher-derivative gravity model, and some quantum corrected BH \cite{Berti:2003zu,Konoplya:2022hll,Konoplya:2022pbc,Konoplya:2022iyn,Konoplya:2023ppx,Konoplya:2023ahd,Fu:2023drp}.

\begin{figure}[htbp]
	\centering
	\includegraphics[width=0.32\textwidth]{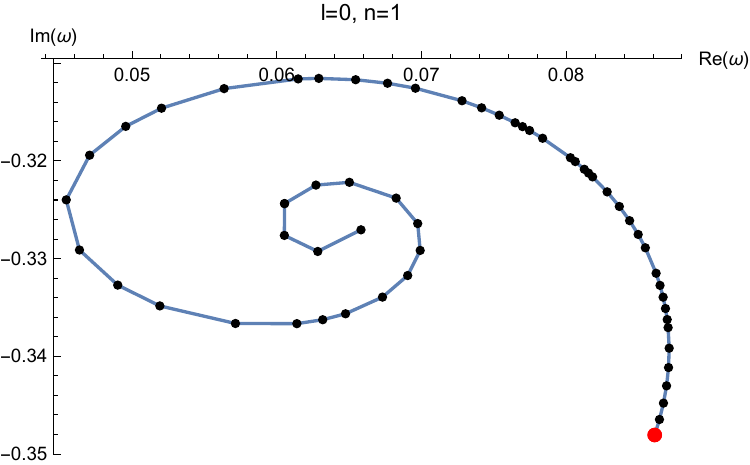}\hspace{1mm}
	\includegraphics[width=0.32\textwidth]{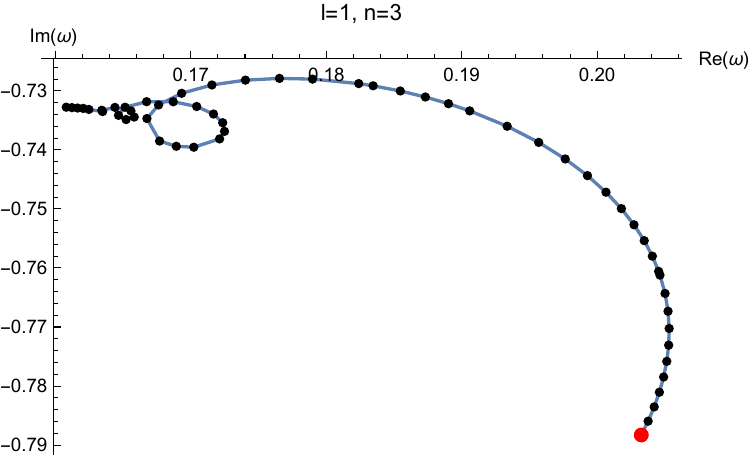}\hspace{1mm}
	\includegraphics[width=0.32\textwidth]{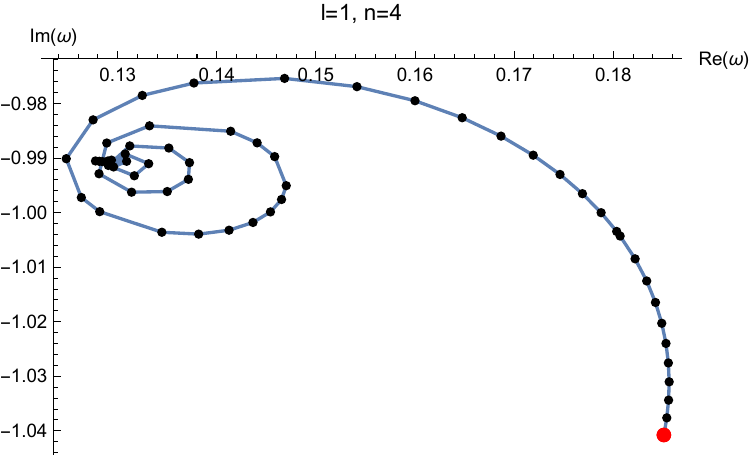}\hspace{1mm}	
	\caption{Phase diagram $Re(\omega)$-$Im(\omega)$ for $(l,n)=(0,1)$, $(l,n)=(1,3)$ and $(l,n)=(1,4)$ from left to right. The red point represent the $\alpha=0$ (Schwarzschild BH)}
	\label{l0}
\end{figure}
We also show the phase diagram $Re(\omega)$-$Im(\omega)$ in Fig.\ref{l0}. The trajectory illustrated in the phase diagram deviates from the Schwarzschild QNMs and spirals toward a stable point. This phenomenon actually reflects the oscillatory behavior previously mentioned and has also been observed in the context of the RN, Kerr and other LQG-corrected BHs \cite{Berti:2003zu,Jing:2008an,Fu:2023drp,Moreira:2023cxy}.

\begin{figure}[htbp]
	\centering
	\includegraphics[width=0.48\textwidth]{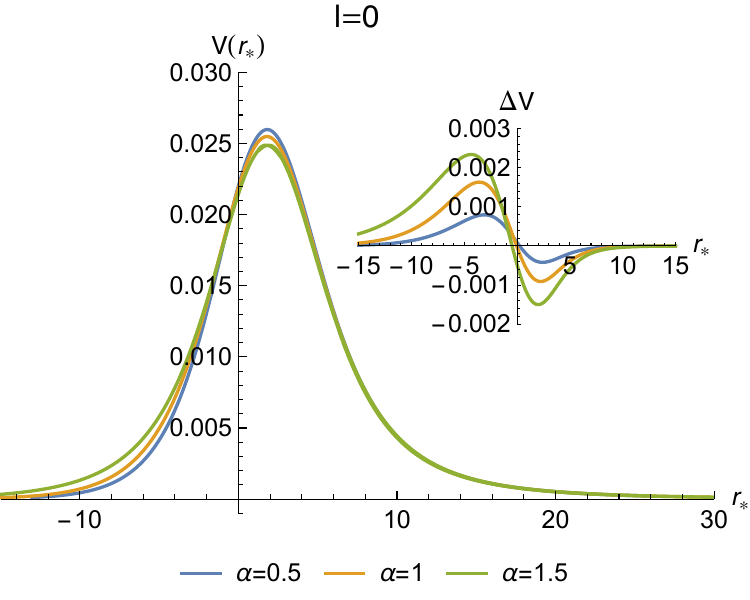}\hspace{5mm}
	\includegraphics[width=0.48\textwidth]{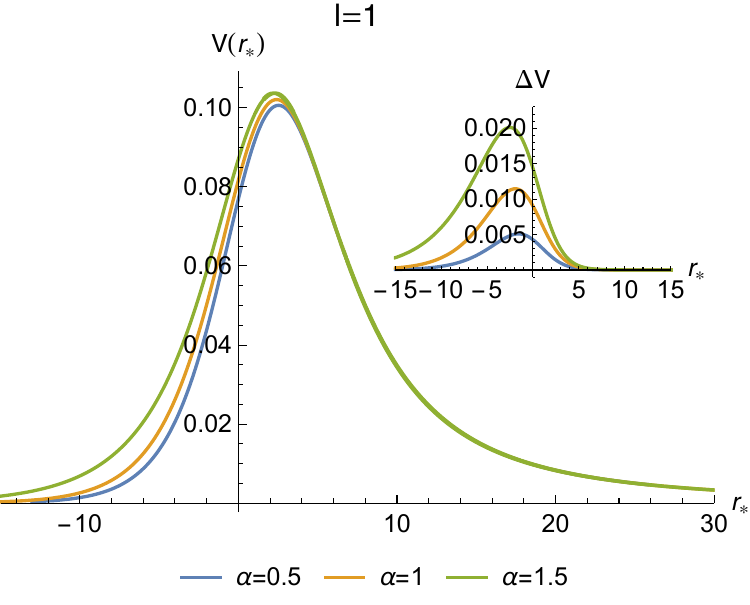}
	\caption{The difference between the effective potential of the LQG-corrected BH and the Schwarzschild BH for different values of $\alpha$ with $l=0$ and $l=1$, denoted as $\Delta V$, respectively.}
	\label{Vvsrsv3}
\end{figure}
The occurrence of this outburst in overtones may be ascribed to disparities within the vicinity of the event horizon between the Schwarzschild BH and the LQG-corrected BH \cite{Berti:2003zu,Konoplya:2022hll,Konoplya:2022pbc,Konoplya:2022iyn,Konoplya:2023ppx,Konoplya:2023ahd,Fu:2023drp}. To this end, we present a comparison of the effective potential between the LQG-corrected and Schwarzschild BHs for different values of $\alpha$ with $l=0$ and $l=1$ (see Fig.\ref{Vvsrsv3}). This comparison is represented by the quantity $\Delta V$, which quantifies the difference in the effective potentials. Evidently, there is a subtle alteration in the vicinity of the event horizon. As we increase the LQG-corrected parameter $\alpha$, the change in the effective potential becomes more pronounced. Such a change in the effective potential may serve as a trigger for the outburst of overtones. This observation aligns with the argument presented in \cite{Konoplya:2022pbc}. Therefore, the quantum gravity effect could be one of the influencing factors resulting in the outburst of the overtones.

\section{Conclusion and discussion}\label{sec-conclusion}

Gravitational wave events provide us with unprecedented opportunities to explore the strong gravitational region and the quantum gravity effect, offering more stringent tests of general relativity. Extracting essential information from gravitational wave signals is paramount, and a key avenue for this endeavor is the investigation of quasi-normal modes in the context of effective quantum gravity BHs as QNMs dominate the ringdown phase of BH mergers. In this paper, we delve into an in-depth investigation of the properties of the fundamental modes of the LQG corrected BH proposed in \cite{Lewandowski:2022zce} and extend our study to include higher overtone cases. The main results are summarized as follows:
\begin{enumerate}
	\item For the fundamental modes, the real part of the QNF for $l=0$ demonstrates a pronounced non-monotonic behavior with respect to the LQG-corrected parameter $\alpha$. However, this non-monotonic behavior disappears for $l>0$. As for the imaginary part, whether for $l=0$ or $l>0$, the QNF exhibits a monotonic behavior. Specifically, we observe the absolute value of the imaginary part decreases with an increase of the LQG-corrected parameter $\alpha$. This indicates that quantum gravity effects lead to slower decay modes.
	\item For higher overtones, a notable deviation becomes evident between the QNFs of the LQG-corrected and Schwarzschild BHs. Specifically, for certain higher overtone modes, this discrepancy can be substantial, reaching magnitudes of several hundred percent, even when the fundamental mode closely aligns with its Schwarzschild counterpart. Consequently, the intervention of quantum gravity corrections leads to a significant outburst of overtones. This outburst of these overtones can be attributed to the distinctions near the event horizons between the Schwarzschild and LQG-corrected BHs. In other words, overtones can serve as a means to probe physical phenomena or disparities in the vicinity of the event horizon.
\end{enumerate}

There is much that warrants further investigation. Here, we discuss some open questions.

\begin{enumerate}
	\item In order to constrain the LQG parameters using gravitational wave data, it is crucial to extend the static LQG-corrected BH to the rotating case. Following this, an investigation into its gravitational perturbations is necessary, along with a further analysis of the the QNM spectrum.
	\item In \cite{Cardoso:2017soq}, the authors point out when the Cauchy and event horizon radius approach each other, a family of near-extreme modes (NE modes) emerges. It is interesting whether the LQG-corrected BH also exhibits these NE modes in extreme cases and to explore how quantum effects affect these NE modes.
	\item The high sensitivity of overtones to small deformations of near-horizon geometry is definitely related to the concept of ``overtone instability", as elucidated in \cite{Jaramillo:2021tmt,Jaramillo:2020tuu}. Therefore, it would be interesting to employ the perturbation method--pseudospectrum--mentioned in \cite{Jaramillo:2021tmt,Jaramillo:2020tuu} to delve deeper into this instability of this LQG-corrected BH model.
	\item The characteristics of QNMs, particularly the fundamental modes, have been extensively studied for various quantum-corrected black holes. Examples include the string-corrected BH \cite{Moura:2021eln,Moura:2021nuh,Moura:2022gqm} and the Schwarzschild BH with generalized uncertainty principle (GUP) \cite{Anacleto:2021qoe}. It would be intriguing to delve into the properties of higher overtones and compare them with the results of our model investigated in this study.
\end{enumerate}

\section*{Acknowledgments}

We are very grateful to Xi-Jing Wang for helpful discussions and suggestions. This work is supported by National Key R$\&$D Program of China (No. 2020YFC2201400), the Natural Science Foundation of China under Grants No. 12375055 and No. 12347159, the Postgraduate Research $\&$ Practice Innovation Program of Jiangsu Province under Grant No.KYCX22$\_$3451, and the Postgraduate Scientific Research Innovation Project of Hunan Province under Grant No. CX20220509.

\appendix
\section{The methods and errors analysis}\label{sec-methods}

In this appendix, we provide a concise overview of the numerical methods utilized in our study and conduct an error analysis.

\subsection{WKB method}\label{sec-WKB}

The WKB method, initially applied to the problem of scattering around BHs \cite{Schutz:1985km}, has undergone significant development from 1st to 13th order to determine the QNFs \cite{PhysRevD.35.3621,Guinn:1989bn,Konoplya:2004ip,Konoplya:2003ii,Matyjasek:2017psv,Konoplya:2019hlu}.
The general higher-order WKB formula is as follows \cite{Konoplya:2019hlu}:
\begin{eqnarray}\label{WKB}
	\omega^2 = V_0 + A_2(\mathcal{K}^2) + A_4(\mathcal{K}^2) + \ldots - i\mathcal{K}\sqrt{-2V_2}\Big(1 + A_3(\mathcal{K}^2) + A_5(\mathcal{K}^2) + \ldots\Big).
\end{eqnarray}
Here, $\mathcal{K}$ takes a half-integer value, and $A_k(\mathcal{K}^2)$ represents the $k$th order correction term, which consists of polynomials of $\mathcal{K}^2$ with rational coefficients. These correction terms depend on the derivative of the effective potential at its maximum position. In this context, $V_i$ denotes the derivative with respect to $r$ at its maximum position, while $V_0$ corresponds to the potential itself at its maximum position.

\begin{figure}[H]
	\centering
	\includegraphics[width=0.48\textwidth]{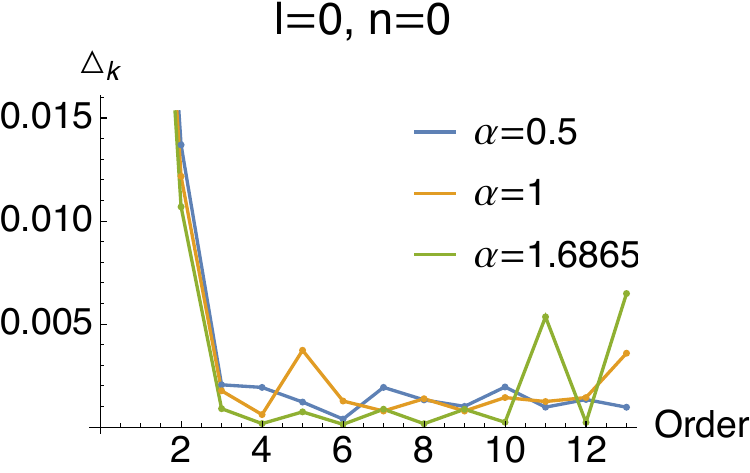}\hspace{1mm}
	\includegraphics[width=0.48\textwidth]{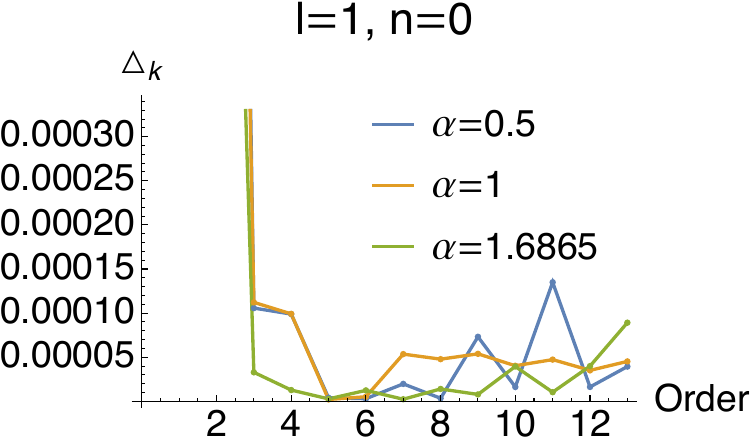}\ \\	
	\includegraphics[width=0.48\textwidth]{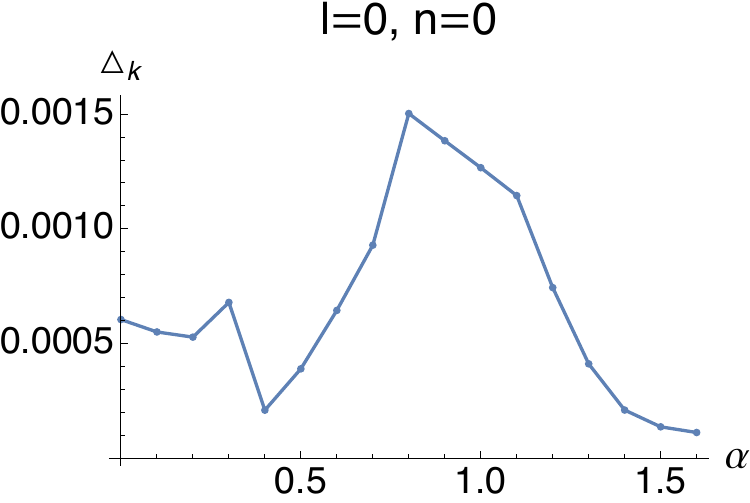}\hspace{1mm}		
	\includegraphics[width=0.48\textwidth]{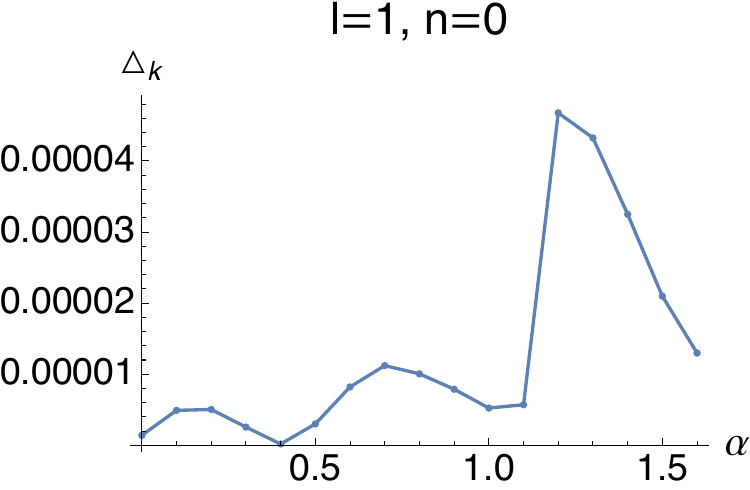}	
	\caption{The plots above: the error estimation function $\Delta_{k}$ as a function of WKB order for the fundamental mode, along with various quantum-corrected parameters $\alpha$. The plots below: $\Delta_{k}$ as a function of $\alpha$ for the sixth-order WKB approximation.}
	\label{ErrorWKB}
\end{figure}
We define the error estimation function $\Delta_{k}$ as \cite{Konoplya:2019hlu} 
\begin{eqnarray}\label{estimation-function}
	\Delta_k=\frac{\left|\omega_{k+1}-\omega_{k-1}\right|}{2}.
\end{eqnarray}
The plots above in Fig.\ref{ErrorWKB} depict the error estimation function $\Delta_{k}$ as a function of WKB order for the fundamental mode, along with various quantum-corrected parameters $\alpha$. Our analysis reveals that the sixth-order WKB approximation yields a relatively small $\Delta_{k}$ in the model. Additionally, we present $\Delta_{k}$ as a function of $\alpha$ for the sixth-order WKB approximation, clearly demonstrating that it remains below $0.15\%$. Based on these findings, we conclude that the sixth-order WKB approximation offers a higher level of accuracy compared to other order WKB approximations in our model. Consequently, we will employ the 6th-order WKB method to determine the QNFs here.

It is essential to recognize that the WKB series converges only asymptotically and does not ensure improved accuracy at each order. Therefore, cross-validating the WKB results with other methods, such as the AIM or the pseudospectral method, becomes necessary. Furthermore, the WKB method inherently fails to capture high overtone modes. As a result, in this study, we primarily utilize the WKB method as an additional check and solely for calculating the fundamental modes. However, it is worth noting that the WKB method offers the advantage of being time-efficient in determining the QNFs.

\subsection{Pseudospectral method}\label{sec-PS}

The pseudospectral method is a fully numerical method and serves as a powerful tool in determining QNFs \cite{Jansen:2017oag,Wu:2018vlj,Fu:2018yqx,Xiong:2021cth,Liu:2021fzr,Liu:2021zmi,Jaramillo:2020tuu,Jaramillo:2021tmt,Destounis:2021lum,Fu:2022cul,Fu:2023drp}, particularly for finding high overtone modes \cite{Fu:2023drp}.
The key point of the pseudospectral method is to discretize the differential equations and then solve the resulting generalized eigenvalue equations. Specially, we replace the continuous variables by a discrete set of collocation points called the grid points and expand the functions by some particular basis functions called cardinal functions.
Usually, we use the Chebyshev grids and Lagrange cardinal functions:
\begin{eqnarray}
	\label{Cg-Lcf}
	x_i=\cos\left( \frac{i}{N}\pi\right) \,, \ \ C_j(x)=\prod_{i=0,i\neq j}^N \frac{x-x_i}{x_j-x_i}\,, \ i=0\,, ...\,, N\,.
\end{eqnarray}

When employing the pseudospectral method to determine the QNFs, we will work in the Eddington-Finkelstein coordinates, where Eq.\eqref{wavelike-equation} is linear in the frequency $\omega$. Specifically, in the Eddington-Finkelstein coordinates, imposing boundary conditions becomes more convenient. Furthermore, we introduce the transforms $r\rightarrow1/u$. After applying this operation, the metric (\ref{metric}) takes the following form:
\begin{eqnarray}\label{metricEF}
	ds^2=-f(u)dt^2+2dtdu+(spatial\;part)\,,
\end{eqnarray}
where
\begin{eqnarray}\label{LQG-BH}
	&&
	f(u)=1-2u+\frac{u^4(2 u_h-1)}{u_h^4}.\
\end{eqnarray}
Here, $u=0$ and $u=u_h$ correspond to the boundary and the horizon, respectively. The parameters $u_h$ and $\alpha$ are related as follows:
\begin{eqnarray}\label{uh-alpha}
	1-2 u_h+u_h^4 \alpha=0\,.
\end{eqnarray}
In this coordinate system, the wave equation (\ref{scalar-eqution2}) becomes:
\begin{eqnarray}\label{wave-eqution}
	\psi''(u)=\lambda_0(u)\psi'(u)+s_0(u)\psi(u),
\end{eqnarray}
where prime denotes the derivative with respect to $u$. The coefficients $\lambda_0(u)$ and $s_0(u)$ are as follows:
\begin{gather}\label{lambda-s}
	\begin{split}
		\lambda_0(u)=&\frac{2 i (-2 u^4 u_h (-2\omega + u (2 \omega +i))-2 u^4 \omega +u^2 u_h^2 (u^3 (8 \omega +4 i)-8 \omega -i)+\omega)}{u^2 u_h (-u^4+2(u^3-1) u u_h+1)}, \\
		s_0(u)=&\frac{1}{u^3 u_h^2 (-u^4+2 (u^3-1) u u_h+1)}(l^2 u u_h^2+l u u_h^2+2 \omega  (-2 u^3\omega -2 u_h^2 (u^5 (4\omega +3 i) \\
		&	-2 u^4 (4 \omega +i)+4 u\omega)+4 u_h^3 (-4u^2 \omega +u^5 (4 \omega +3i))-u_h (u^4 (8\omega +2 i)-4 u^3 \omega + i))).
	\end{split}
\end{gather}
Afterward, we can obtain the generalized eigenvalue equation in the form:
\begin{eqnarray}\label{eq1}
	(M_0+\omega M_1)\psi=0,
\end{eqnarray}
where $M_i$ ($i=0,1$) represents the linear combination of the derivative matrices. By solving the eigenvalue function directly, we can determine the QNFs.

As a fully numerical method, the pseudospectral method is a powerful tool for finding QNFs. This method can accurately solve differential equations and operates in the frequency domain, making it highly efficient in computing QNFs. However, like many numerical methods, one of the drawbacks of the pseudospectral method is its potential to be time-consuming when compared to semi-analytic methods such as the WKB method and the AIM. For complex problems or large-scale models, calculating QNFs may require considerable time, particularly as the number of spatial dimensions and sampling points increases. Consequently, when using the pseudospectral method, careful consideration is necessary to strike a balance between computational time and the desired accuracy.

\subsection{Asymptotic iteration method}\label{sec-AIM}

The AIM can be utilized to solve eigenvalue problems and second-order differential equations \cite{Ciftci:2005xn,Ciftci_2003,Ismail_2020}. The improved version of the AIM can be employed to accurately calculate QNMs \cite{Cho:2009cj,Cho:2011sf}. 

The main essence of the AIM is to leverage the invariant structure found in the right-hand side of the wave equation \eqref{wave-eqution} \cite{Ciftci:2005xn,Ciftci_2003,Ismail_2020}. By differentiating Eq. \eqref{wave-eqution} $n$ times, we obtain the following expression:
\begin{eqnarray}\label{n2equ}
	\psi^{n+2}(u)=\lambda_n(u)\psi'(u)+s_n(u)\psi(u),
\end{eqnarray}
with the recurrence relations:
\begin{eqnarray}
	&\lambda_n(u)=& \lambda'_{n-1}(u)+s_{n-1}(u)+\lambda_0(u)\lambda_{n-1}(u)\,, \nonumber \\
	&s_n(u)=& s'_{n-1}(u)+s_0(u) \lambda_{n-1}(u)\,.
	\label{recurrence}
\end{eqnarray}

For sufficiently large $n$, we can set up the following relation:
\begin{eqnarray}\label{quan-condtion}
	\frac{s_n(u)}{\lambda_n(u)}=\frac{s_{n-1}(u)}{\lambda_{n-1}(u)}\,.
\end{eqnarray}
This is known as the ``quantization condition'', which is equivalent to 
\begin{eqnarray}\label{quan-condtionv1}
	\delta_n=s_n(u)\lambda_{n-1}(u)-s_{n-1}(u)\lambda_{n}(u)=0\,.
\end{eqnarray}
From the above equation, we can determine the QNFs.

Furthermore, an improved version of the AIM has been designed to prevent taking derivatives at each step \cite{Cho:2009cj,Cho:2011sf}. To achieve this, $\lambda_n$ and $s_0$ are expanded in a Taylor series around the point $\bar{u}$, where the AIM is performed:
\begin{eqnarray}\label{Taylor}
	&\lambda_n(u)=&\sum \limits_{j=0}^\infty c_n^j(u-\bar{u})^j\,, \nonumber \\
	&s_n(u)=&\sum \limits_{j=0}^\infty d_n^j(u-\bar{u})^j\,.
\end{eqnarray}
Here, $c_n^j$ and $d_n^j$ represent the $j$th Taylor coefficients of $\lambda_n(u)$ and $s_n(u)$, respectively. Upon substituting these expressions into Eq. \eqref{recurrence}, a set of recursion relations for the coefficients can be derived as:
\begin{eqnarray}\label{coe}
	&c_n^j=&(j+1)c_{n-1}^{j+1}+d_{n-1}^{j}+\sum \limits_{k=0}^j c_0^k c_{n-1}^{j-k}\,, \nonumber \\
	&d_n^j=&(j+1)d_{n-1}^{j+1}+\sum \limits_{k=0}^j d_0^k c_{n-1}^{j-k}\,.
\end{eqnarray}
Using these coefficients, the ``quantization condition'' \eqref{quan-condtionv1} can be rewritten as:
\begin{eqnarray}\label{quan-condtionv2}
	d_n^0c_{n-1}^0-d_{n-1}^0c_n^0=0\,.
\end{eqnarray}
As a result, the AIM is transformed into a set of recursion relations that no longer necessitate the use of the derivative operator, significantly enhancing both the accuracy and efficiency of the method.

Compared to the WKB method, the AIM is capable of calculating relatively high overtones and is particularly well-suited for computing pure imaginary modes \cite{Ponglertsakul:2018rot}. Additionally, when compared to the pseudospectral method, the AIM exhibits significantly faster calculation speeds. Furthermore, in contrast to the CFM, the AIM can be applied independently of the singularity structure of the ordinary differential equation, making it applicable to a broader range of equations. Conversely, the CFM requires a thorough analysis of the regular singular points (RSP) in the equation, involving lengthy calculations to prepare the recurrence relation coefficients, which are prone to errors during manipulation \cite{Cho:2011yp}.

However, we wish to emphasize that while increasing the number of iterations in the AIM may be necessary as overtones are raised in general \cite{Zhang:2015jda}, more iterations do not necessarily guarantee better results. The optimal number of iterations and unfolding points may vary depending on the specific BH parameters, and their determination often relies on empirical judgment or comparison with other methods. Additionally, the choice of the Taylor expansion point $\bar{u}$ and the Taylor order $j$ can significantly impact the convergence speed of the AIM, as demonstrated in Ref. \cite{Cho:2009wf}. This outcome hinges on the experience and continuous debugging of the approach. 

\subsection{Error analysis}

In this subsection, we will compare the results of the QNFs obtained using the 6th order WKB method or the AIM  with those obtained using the pseudospectral method. Notice that the relative error is computed under the assumption that the values obtained from the pseudospectral method are accurate. To this end, we define:
\begin{eqnarray}
	\varepsilon_1&=\left |\frac{\omega_{PS}-\omega_{WKB}}{\omega_{PS}}\right | \times 100\%\,, \nonumber \\
	\varepsilon_2&=\left |\frac{\omega_{PS}-\omega_{AIM}}{\omega_{PS}}\right | \times 100\%\,.
	\label{error}
\end{eqnarray}
We will use $\varepsilon^{Re}_{\#}$ and $\varepsilon^{Im}_{\#}$ to represent the errors in the real and imaginary parts of the QNFs, respectively.

Table \ref{M1l0n0} displays the QNFs for $l=0$ and $n=0$, which are calculated using different methods for various $\alpha$ values. The corresponding errors of the real and imaginary parts, $\varepsilon^{Re}_{\#}$ and $\varepsilon^{Im}_{\#}$, are presented in Table \ref{M1l0n0v1}. It can be observed that the 6th order WKB method exhibits relatively large errors compared to the pseudospectral method. However, the results obtained from the AIM closely match those from the pseudospectral method.
\begin{table}[H]
	\setlength\tabcolsep{10pt}
	\begin{center}
		\caption{QNFs are calculated using different methods for various $\alpha$ values. In this context, ``PS'' refers to the pseudospectral method, and we specifically focus on the case where $l=0, n=0$.}
		\begin{tabular}{|c|c|c|c|}
			\hline
			$\alpha$  & \text{$\omega $(WKB)} & \text{$\omega $(PS)} & \text{$\omega
				$(AIM)} \\
			\hline
			0 & 0.11046702 - 0.10081625$i$ & 0.11045494 - 0.10489572$i$ &
			0.11045500 - 0.10489571$i$  \\
			\hline
			0.2500 & 0.11177388 - 0.10109890$i$  & 0.11143219 - 0.10361592$i$  &
			0.11143224 - 0.10361592$i$  \\
			\hline
			0.5000 & 0.11477595 - 0.09866554$i$  & 0.11238550 - 0.10207584$i$  &
			0.11238554 - 0.10207585$i$  \\
			\hline
			0.7500 & 0.11979027 - 0.09115157$i$  & 0.11325044 - 0.10018558$i$  &
			0.11325047 - 0.10018560$i$  \\
			\hline
			1.0000 & 0.12319526 - 0.08134095$i$  & 0.11387739 - 0.09781173$i$  &
			0.11387742 - 0.09781175$i$  \\
			\hline
			1.5000 & 0.11467730 - 0.07528910$i$ & 0.11191203 - 0.09141316$i$  &
			0.11191397 - 0.09141256$i$ \\
			\hline
			1.6000 & 0.11375856 - 0.07531746$i$ & 0.11026897 - 0.09109497$i$  &
			0.11040659 - 0.09104516$i$  \\
			\hline
			1.6500 & 0.11346383 - 0.07523894$i$ & 0.10992184 - 0.09129992$i$  &
			0.10992206 - 0.09130020$i$ \\
			\hline
			1.6865 & 0.11326678 - 0.07515483$i$  & 0.10972530 - 0.09127475$i$ &
			0.10972599 - 0.09127493$i$  \nonumber\\
			\hline
		\end{tabular} \\
		\label{M1l0n0}
	\end{center}
\end{table}
\begin{table}[H]
	\setlength\tabcolsep{3pt}
	\begin{center}
		\caption{The errors of the 6th order WKB method and the AIM compared with those of the pseudospectral  method for various $\alpha$ values, corresponding to those in Table \ref{M1l0n0}. An error of $0\%$ indicates that the difference is smaller than $10^{-7}$. Here, $l=0, n=0$.}
		\begin{tabular}{|c|c|c|c|c|c|c|c|c|c|c|c|c|}
			\hline
			$\alpha$ & 0  & 0.2500 & 0.5000 & 0.7500 & 1.0000 & 1.5000 & 1.6000 & 1.6500 & 1.6865 \\
			\hline
			$\varepsilon^{Re}_{1}$ & 0.01094\%  & 0.30664\% & 2.12701\% & 5.77466\% & 8.18237\% & 2.47093\% & 3.16462\% & 3.22229\% & 3.22759\% \\
			\hline
			$\varepsilon^{Im}_{1}$ & 3.88907\% & 2.42918\% & 3.34095\% & 9.01728\% & 20.2221\% & 17.6387\% & 17.3199\% & 17.5914\% & 17.6609\% \\
			\hline
			$\varepsilon^{Re}_{2}$ & 0\% & 0\% & 0\% & 0\% & 0\% & 0.00173\% & 0.12480\% & 0.00020\% & 0.00063\% \\
			\hline
			$\varepsilon^{Im}_{2}$ & 0\% & 0\% & 0\% & 0\% & 0\% & 0.00066\% & 0.05468\% & 0.00031\% & 0.00020\% \\
			\hline
		\end{tabular}
		\label{M1l0n0v1}
	\end{center}
\end{table}
\begin{table}[H]
	\setlength\tabcolsep{10pt}
	\begin{center}
		\caption{QNFs are calculated using different methods for high overtones. Here, we fix $l=0$ and $\alpha=0.5$.}
		\begin{tabular}{|c|c|c|c|}
			\hline
			\text{n} & \text{$\omega $(WKB)} & \text{$\omega $(PS)} &
			\text{$\omega $(AIM)} \\
			\hline
			0 & 0.1147760 - 0.0986655$i$ & 0.1123855 - 0.1020758$i$ &
			0.1123855 - 0.1020759$i$ \\
			\hline
			1 & 0.0934456 - 0.3271466$i$ & 0.0870026 - 0.3365118$i$ &
			0.0871007 - 0.3365497$i$ \\
			\hline
			2 & 0.1746372 - 0.4548715$i$ & 0.0713298 - 0.5813296$i$  &
			0.0722420 - 0.5805860$i$  \\
			\hline
			3 & 0.9197044 - 0.3375424$i$ & 0.0572010 - 0.8276979$i$ &
			0.0588690 - 0.8305190$i$ \\
			\hline
			4 & 2.4581780 - 0.4374126$i$ & 0.0415393 - 1.0717112$i$ &
			0.0454604 - 1.0688645$i$ \\
			\hline
			5 & 4.8159854 - 0.6238195$i$ & 0.0357523 - 2.1360647$i$ &
			0.0413650 - 1.8667191$i$ \nonumber\\
			\hline
		\end{tabular} \\
		\label{M1a0p5l0}
	\end{center}
\end{table}
\begin{table}[H]
	\setlength\tabcolsep{8pt}
	\begin{center}
		\caption{The errors of the 6th order WKB method and the AIM compared with those of the pseudospectral  method for high overtones. Here, we fix $l=0$ and $\alpha=0.5$.}
		\begin{tabular}{|c|c|c|c|c|c|c|c|c|c|c|c|c|}
			\hline
			$n$ & 0  & 1 & 2 & 3 & 4 & 5  \\
			\hline
			$\varepsilon^{Re}_{1}$ & 2.12701\%  & 7.40550\% & 144.831\% & 1507.85\% & 5817.71\% & 13370.4\%  \\
			\hline
			$\varepsilon^{Im}_{1}$ & 3.34095\% & 2.78302\% & 21.7533\% & 59.2191\% & 59.1856\% & 70.7959\%   \\
			\hline
			$\varepsilon^{Re}_{2}$ & 0.00004\% & 0.11280\% & 1.27889\% & 2.91601\% &  9.43944\% & 15.6988\%  \\
			\hline
			$\varepsilon^{Im}_{2}$ & 0.00001\% & 0.01127\% & 0.12791\% & 0.34084\% & 0.26562\% & 12.6094\%  \\
			\hline
		\end{tabular}
		\label{M1a0p5l0v1}
	\end{center}
\end{table}
\begin{table}[H]
	\setlength\tabcolsep{8pt}
	\begin{center}
		\caption{QNFs are calculated using different methods for high overtones. Here, we fix $l=1$ and $\alpha=0.5$.}
		\begin{tabular}{|c|c|c|c|}
			\hline
			\text{n} & \text{$\omega $(WKB)} & \text{$\omega $(PS)} &
			\text{$\omega $(AIM)} \\
			\hline
			0 & 0.2960292440 - 0.0955759232$i$  & 0.2960395851 - 0.0953703511$i$&
			0.2960395851 - 0.0953703511$i$  \\
			\hline
			1 & 0.2693777655 - 0.2983601907$i$  & 0.2690310583 - 0.2977723864$i$ &
			0.2690310582 - 0.2977723866$i$  \\
			\hline
			2 & 0.2364134322 - 0.5244490994$i$ & 0.2340497112 - 0.5227186265$i$&
			0.2340496098 - 0.5227186825$i$ \\
			\hline
			3 & 0.2226805321 - 0.7650800812$i$ & 0.2046358065 - 0.7612060062$i$&
			0.2046343470 - 0.7612001844$i$\\
			\hline
			4 & 0.2569610993 - 0.9964677978$i$ & 0.1806951756 - 1.0042841446$i$&
			0.1806936202 - 1.0043494301$i$ \\
			\hline
			5 & 0.3833521733 - 1.1817052053$i$& 0.1592163047 - 1.2487030878$i$&
			0.1595266385 - 1.2487265580$i$ \nonumber\\
			\hline
		\end{tabular}
		\label{M1a0p5l1}
	\end{center}
\end{table}
\begin{table}[H]
	\setlength\tabcolsep{10pt}
	\begin{center}
		\caption{The errors of the 6th order WKB method and the AIM compared with those of the pseudospectral  method for high overtones. Here, we fix $l=1$ and $\alpha=0.5$.}
		\begin{tabular}{|c|c|c|c|c|c|c|c|c|c|c|c|c|}
			\hline
			$n$ & 0  & 1 & 2 & 3 & 4 & 5  \\
			\hline
			$\varepsilon^{Re}_{1}$ & 0.00349\%  & 0.12887\% & 1.00992\% & 8.81797\% & 42.2070\% & 140.774\%  \\
			\hline
			$\varepsilon^{Im}_{1}$ & 0.21555\% & 0.19740\% & 0.33105\% & 0.50894\% & 0.77830\% & 5.36540\%   \\
			\hline
			$\varepsilon^{Re}_{2}$ & 0\% & 0\% & 0.00004\% & 0.00071\% &  0.00086\% & 0.19491\%  \\
			\hline
			$\varepsilon^{Im}_{2}$ & 0\% & 0\% & 0.00001\% & 0.00076\% & 0.00650\% & 0.00188\%  \\
			\hline
		\end{tabular}
		\label{M1a0p5l1v1}
	\end{center}
\end{table}

Then, we examine the scenario with high overtones. Tables \ref{M1a0p5l0} and \ref{M1a0p5l0v1} present the QNFs calculated using different methods for high overtones, with $l=0$ and $\alpha=0.5$. Additionally, Tables \ref{M1a0p5l1} and \ref{M1a0p5l1v1} showcase the results for $l=1$ and $\alpha=0.5$. It is evident that the WKB method completely loses its power to determined the high overtone modes.
With the AIM, the errors are relatively large for $l=0$ with high overtones, but they remain within a tolerable range for $l=1$, even for $n=5$. However, at this point, the AIM requires at least $100$ iterations.

Based on the aforementioned observations and discussions, in this paper, we primarily utilize the pseudospectral method to determine the QNFs. The WKB method and the AIM are employed solely for crosschecking purposes.

\bibliographystyle{style1}
\bibliography{Ref}
\end{document}